\documentclass[12pt]{article}
\def\lae{\;^{<}_{\sim} \;} 

\title{\textbf{Large non-Gaussianity generated after D-term Inflation
by Right-Handed Sneutrino Curvaton }}
{\author{\\[1cm]
{\sc \large Chia-Min Lin$^{1}$ and Kingman Cheung$^{2}$}\\
{\sl\small Department of Physics, National Tsing Hua University, Hsinchu, Taiwan 300 }\\
{\sl\small Physics Division, National Center for Theoretical Sciences,
Hsinchu 300, Taiwan}\\
{\sl\small Division of Quantum Phases \& Devices, School of Physics,
Konkuk university, Seoul 143-701, Korea}
}}

\usepackage[margin=2cm]{geometry}
\usepackage{graphicx,color}
\usepackage{subfigure}
\begin{document}
\maketitle
\begin{abstract}
 In this paper, we consider large non-Gaussianity generated after
 D-term inflation in the case that the adiabatic curvature
 perturbation produced is dominated by a right-handed (RH) sneutrino
 curvaton. The cosmic string problem can also be evaded with
around $10\%$
 contribution to the CMB power spectrum and the spectral index $n_s
 \simeq 1$. The non-Gaussianity produced can be as large as
 $10<f_{NL}<100$ for the right-handed sneutrino mass being around
 $m_\Phi=100\mbox{ GeV}$ with the Yukawa coupling $\lambda_\nu \sim
 10^{-10}$. In this case, the lightest neutrino mass is
 $10^{-10}\mbox{ eV}$ which favors RH sneutrino leptogenesis.
\end{abstract}
\footnoterule{\small $^1$cmlin@phys.nthu.edu.tw, $^2$cheung@phys.nthu.edu.tw}
\section{Introduction}
Inflation \cite{Starobinsky:1980te, Sato:1980yn, Guth:1980zm} (for
review, \cite{Lyth:1998xn, Lyth:2007qh, Linde:2007fr}) is a vacuum
energy dominated epoch in the early universe when the scale factor
grew quasi-exponentially. This scenario is used to set the initial
condition for the subsequent hot big bang and provides primordial
density perturbation as the seed of structure formation. There are a
lot of inflation models being introduced in literature. The spectral
index $n_s$ is a useful observable to discriminate among various
inflation models.  However, as experiments improve, there exists
another useful observable -- non-Gaussianity -- characterized in
some models with a non-linearity parameter $f_{NL}$. For a review of
non-Gaussianity see Ref.~\cite{Bartolo:2004if}.


The non-linearity parameter $f_{NL}$ takes the form
\begin{equation}
\zeta(x)=\zeta_g(x)+\frac{3}{5}f_{NL}\zeta^2_g(x)+\cdots,
\end{equation}
where $\zeta$ is the curvature perturbation in the uniform density
slice, $\zeta_g$ denotes the Gaussian part of $\zeta$. The
non-linearity parameter $f_{NL}$ parameterizes the non-Gaussianity
forming the irreducible 3-point correlation function.

Large primordial (local-type) non-Gaussianity was reported
($2$-$\sigma$ range) \cite{Yadav:2007yy}
\begin{equation}
27 < f_{NL}<147 \;,
\end{equation}
while the result from WMAP team is ($2$-$\sigma$ range)  \cite{Komatsu:2008hk}
\begin{equation}
-9<f_{NL}<111 \;,
\end{equation}
which is more conservative.
Recently, a new results ($2$-$\sigma$ range)
\begin{equation}
-18<f_{NL}<80
\end{equation}
is obtained in \cite{Curto:2009pv}.  In the near future, the Planck
satellite \cite{:2006uk} will reduce the bound to $|f_{NL}| \lae 10$
if non-Gaussianity is not detected. Therefore, we investigate the
possibility of producing $10<f_{NL}<100$ in this work.  We refer
this range as large non-Gaussianity.

In the framework of supersymmetric inflation models, one of the
interesting models is D-term hybrid inflation. It is well-known that
hybrid inflation alone \footnote{It is possible to produce large
  non-Gaussianity after hybrid inflation via Tachyonic preheating as in
  \cite{Barnaby:2006km}. We don't consider this possibility in this
  paper though.} produces very small non-Gaussianity ($|f_{NL}|<1$)
\cite{Vaihkonen:2005hk}. If large non-Gaussianity is detected in the
future, the model will face great difficulty. This issue was
mentioned in \cite{Bernardeau:2007xi}, which pointed out that D-term
inflation has to be extended in order to generate detectable
non-Gaussianity.  In this work, we investigate the possibility of
producing large non-Gaussianity in D-term inflation by adding a
right-handed sneutrino field, which plays the role of a curvaton. The
possibility of using right-handed(RH) sneutrino as a curvaton was
considered in \cite{McDonald:2003xq, McDonald:2004by, Moroi:2002vx,
  Postma:2002et}. However, non-Gaussianity was not a big issue at that
time, and so the parameter space that can generate large
non-Gaussianity was not explored.

The idea of introducing a right-handed sneutrino field into D-term
inflation has been considered by one of us in Refs.~\cite{Lin:2007va,
  Lin:2006xta}. The purpose was to transform the potential into a
hilltop form \cite{Kohri:2007gq} so as to reduce both the spectral
index and the cosmic string energy per unit length via supergravity
effects. The required mass of the right-handed sneutrino is about
$10^{11}\mbox{ GeV}$. On the other hand, we show that if the
right-handed sneutrino mass is much smaller, around $100\mbox{
GeV}$, it can play the role of the curvaton, generate large
non-Gaussianity and also suitable for leptogenesis.

The organization of this paper is as follows. In Sec. \ref{2}, we
summarize the formalism and describe non-Gaussianity generated in the
curvaton scenario. In Sec. \ref{3}, we introduce a
right-handed sneutrino field,  which plays the role of a curvaton. In
Sec. \ref{4}, we apply this right-handed sneutrino curvaton to
D-term hybrid inflation and show its consequence. Section \ref{5} is
our conclusion.

\section{Non-Gaussianity in curvaton scenario}
\label{2}
The curvaton \cite{Enqvist:2001zp, Lyth:2001nq, Moroi:2001ct,
  Lyth:2002my} is another source of fluctuations other than the
inflaton which can contribute to the cosmic density
fluctuations. During inflation, the energy density is dominated by the
potential of the inflaton, thus the fluctuations of the curvaton field
are initially of isocurvature type. After inflation, the inflaton
energy density is converted into radiations. If the curvaton field
decays later than the inflaton field, the density fraction of the
curvaton to the total energy grows and eventually generates the
adiabatic curvature perturbation after its decay. Thus, in this case
constraints on models of inflation can be alleviated to some
extent. It is also shown that large (local type) non-Gaussianity can
be produced in this scenario \cite{Bartolo:2003jx, Malik:2006pm,
  Sasaki:2006kq}.

Here we give a brief review of the non-Gaussianity generated from
curvaton.  We consider a potential of curvaton $\sigma$ as follows (we
follow the notation used in \cite{Huang:2008ze})
\begin{equation}
V(\sigma)=\frac{1}{2}m^2\sigma^2.
\end{equation}
During inflation we assume $m \ll H$, $\sigma$ is therefore
slow-rolling, which means the field value can be taken as a constant
during inflation.

The amplitude of quantum fluctuation of the curvaton field in a
quasi-de Sitter space is given by
\begin{equation}
\delta \sigma=\frac{H_\ast}{2\pi} \;,
\end{equation}
where $\ast$ denotes the epoch of horizon exit during inflation.
The spectrum of fractional perturbation is
\begin{equation}
P^{1/2}_{\delta \sigma/ \sigma} \simeq \frac{\delta \sigma}{\sigma}
=\frac{1}{2\pi}\frac{H_\ast}{\sigma_\ast}.
\end{equation}
Initially since the curvaton energy density is subdominant, its
fluctuations are of isocurvature type. After inflation and curvaton
decay, the initial isocurvature perturbation is converted into
adiabatic curvature perturbation which is given by \cite{Lyth:2002my}
\begin{equation}
P^{1/2}_{\zeta_\sigma}=\frac{2}{3}\Omega_{\sigma,D}P^{1/2}_{\delta \sigma/ \sigma}
=\frac{1}{3}\Omega_{\sigma,D}\frac{H_\ast}{\sigma_\ast} \;,
\label{eq2}
\end{equation}
where
\begin{equation}
\Omega_{\sigma,D} \equiv \left(\frac{\rho_\sigma}{\rho_{tot}}\right)_D \;.
\end{equation}
If the curvature perturbation is dominated by that generated from
curvaton, the spectral index takes the form
\begin{equation}
n_s=1+2\eta_{\sigma\sigma}-2\epsilon_H \;,
\label{eq3}
\end{equation}
where
\begin{equation}
\eta_{\sigma\sigma} \equiv \frac{1}{3H^2}\frac{d^2V(\sigma)}{d\sigma^2} \;\;\;
\mbox{and}\;\;\; \epsilon_H \equiv -\frac{\dot{H}_\ast}{H^2_\ast}.
\end{equation}
The bispectrum is given by \cite{Lyth:2002my, Lyth:2006gd}
\begin{equation}
f_{NL} \simeq \frac{5}{4 \Omega_{\sigma,D}}.
\label{eq6}
\end{equation}
This shows that the non-Gaussianity increases as the curvaton energy
density decreases. A large non-Gaussianity is obtained if
$\Omega_{\sigma, D} \ll 1$.

We denote $t_0$ as the time when the curvaton starts to oscillate,
$t_D$ the time when the curvaton decays and assume
$a(t_0)=1$. $\Gamma$ denotes the decay rate of the curvaton. The
energy density of curvaton and radiation are
\begin{eqnarray}
\rho_\sigma(t_0) &=& \frac{1}{2}m^2\sigma^2_\ast, \\
\rho_R (t_0)     &=& 3M_P^2 m^2,  \\
\rho_R(t_D) &\simeq& 3M_P^2 \Gamma^2 \simeq \rho_R(t_0)a^{-4}(t_D),
\end{eqnarray}
where we use the fact that radiation energy density goes like $a^{-4}$.
Therefore,
\begin{equation}
a(t_D) \simeq \left(\frac{m}{\Gamma}\right)^{1/2} \mbox{ and} \;\;\;\;
\Omega_{\sigma, D} \simeq \frac{\rho_\sigma(t_D)}{\rho_R(t_D)}=
\frac{\sigma^2_\ast}{6 M_P^2} \left(\frac{m}{\Gamma}\right)^{1/2} \;,
\end{equation}
where we have assumed that the radiation dominates when the curvaton decays.
This is valid as long as we consider large non-gaussianity
produced. From Eq. (\ref{eq6}), we have
\begin{equation}
f_{NL}=\frac{15}{2}\frac{M_P^2}{\sigma^2_\ast}\left(\frac{\Gamma}{m}\right)^{1/2}
\;.
\label{eq12}
\end{equation}

If the curvaton dominates the curvature perturbation, then from CMB
normalization ($P_R=2.457 \times 10^{-9}$) \cite{Komatsu:2008hk} and
using Eq. (\ref{eq2}, \ref{eq6}), we obtain
\begin{equation}
f_{NL}=2.68\times 10^3 \frac{H_\ast}{\sigma_\ast} \;.
\label{cmbfnl}
\end{equation}

\section{Right Handed Sneutrino as a Curvaton}
\label{3}

There are evidences that neutrino masses are non-vanishing as indicated
by the neutrino oscillations observed by the Super-Kamiokande
\cite{Fukuda:2000np} and SNO collaborations \cite{Ahmad:2002ka}. The
lightness of the neutrino could be explained by the see-saw mechanism.
The superpotential of the mass eigenstate RH neutrinos, $\Phi_i$, is given by
\begin{equation}
W_\nu=\lambda_\nu\Phi H_u L +\frac{m_\Phi\Phi^2}{2},
\end{equation}
where $\Phi$ is the RH neutrino superfield, $H_u$ and $L_i$ are the
MSSM Higgs and lepton doublet superfields, and $m_{\Phi}$ is the RH neutrino
mass. The first term describes the Yukawa coupling of the RH
neutrino. This leads to Majorana neutrino masses via the see-saw
mechanism \cite{seesaw}. The neutrino mass $m_\nu$ is given by the
eigenvalue of the see-saw mass matrix.
\begin{equation}
m_\nu=\frac{m^2_{\nu D}}{m_\Phi},
\label{eq1}
\end{equation}
where $m_{\nu D}=\lambda_D \langle H_u\rangle$ is the Dirac mass
coming from Higgs mechanism and $\langle H_u\rangle \simeq 174 \mbox{ GeV}$.

The decay rate for right-handed sneutrino is
\begin{equation}
\Gamma_{\Phi} \simeq \frac{\lambda^2_\nu}{4 \pi}m_\Phi \;.
\label{decay}
\end{equation}
We are going to use RH sneutrino as the curvaton field, hence from
Eq. (\ref{eq12}, \ref{cmbfnl}), we obtain
\begin{equation}
f_{NL}=\frac{15}{2}\frac{M_P^2}{\Phi^2_\ast}\frac{\lambda_\nu}{\sqrt{4\pi}}
=2.68\times10^3\frac{H_\ast}{\Phi_\ast}
\label{eq15}
\end{equation}
The validity of this equation is based on the assumption that the
classical field value $\Phi_\ast$ is larger than its quantum
fluctuation $H/2\pi$ during inflation. Therefore, besides constraint
from observation ($f_{NL} \lae 100$), there is an upper bound for
the possible non-Gaussianity ($f_{NL} \lae 1000$). We cannot simply
choose small $\Phi_\ast$ to get arbitrary large non-Gaussianity.
\section{D-term Inflation and Non-gaussianity}
\label{4}
The superpotential of D-term hybrid inflation is given by
\cite{Binetruy:1996xj,Halyo:1996pp, Riotto:1997wy, Lyth:1997pf}
\begin{equation}
W=\lambda S \Phi_+ \Phi_-  \;,
\end{equation}
where $S$ is the inflaton superfield, $\lambda$ is the
superpotential coupling, and $\Phi_\pm$ are chiral superfields
charged under the $U(1)_{FI}$ gauge symmetry responsible for the
Fayet-Iliopoulos term. The corresponding scalar potential is
\begin{equation}
V(S, \Phi_+, \Phi_-)=\lambda^2\left[|S|^2(|\Phi_+|^2
+|\Phi_-|^2)+|\Phi_+|^2|\Phi_-|^2\right]+\frac{g^2}{2}
\left(|\Phi_+|^2-|\Phi_-|^2+\xi\right)^2 \;,
\end{equation}
where $\xi$ is the Fayet-Iliopoulos term and $g$ is the $U(1)_{FI}$
gauge coupling. A very small $g$ (far smaller than order $O(1)$) is
regarded as unnatural, because we do not know of any small (for
example $g<10^{-3}$) gauge couplings in particle physics. The true
vacuum of the potential is given by $|S|=|\Phi_+|=0$ and
$|\Phi_-|=\sqrt{\xi}$.  When $|S| > |S_c|= g \xi^{1/2}/\lambda$,
there is a local minimum occurred at $|\Phi_+|=|\Phi_-|=0$.
Therefore,  at tree level the potential is just a constant $V_0=g^2
\xi^2/2$. The 1-loop corrections to $V$ can be calculated using the
Coleman-Weinberg formula \cite{Coleman:1973jx}
\begin{equation}
\Delta V=\frac{1}{64\pi^2}\sum_i(-1)^F m^4_i\ln\frac{m^2_i}{\Lambda^2},
\end{equation}
with $m_i$ being the mass of a given particle, where the sum goes
over all particles with $F=0$ for bosons and $F=1$ for fermions and
$\Lambda$ is a renormalization scale. Thus the 1-loop potential is
given by (setting $\phi=\sqrt{2}\mbox{Re}(S)$)
\begin{equation}
V(S)=V_0\left(1+\frac{g^2}{4\pi^2}\ln\left(\frac{\phi}{\Lambda}\right)\right),
\end{equation}
where $V_0=g^2\xi^2/2$.

The curvature perturbation generated from inflaton field is
\begin{equation}
P_\zeta=\frac{1}{12\pi^2M_P^6}\frac{V^3}{V^{\prime 2}} \;,
\end{equation}
where $V^\prime \equiv \partial V/\partial \phi$. Hence the
curvature perturbation in D-term inflation is given by
\begin{equation}
P_\zeta=\frac{\xi^2 N}{3 M_P^4} \;.
\label{6}
\end{equation}
In this case the observed curvature perturbation is obtained when
\begin{equation}
\xi^{1/2}=7.9 \times 10^{15} \left(\frac{60}{N}\right)^{1/4}\mbox{ GeV}\;.
\end{equation}

After inflation the $U(1)_{FI}$ gauge symmetry is spontaneously
broken by the VEV of $\Phi_-$, cosmic strings form. The mass per
unit length of the string is given by
\begin{equation}
\mu=2 \pi \xi \;.
\label{dstring}
\end{equation}
The $10\%$ cosmic string bound corresponds to an upper bound
for $\xi$ \cite{Bevis:2006mj, Lin:2008ys}:
\begin{equation}
\xi^{1/2} \lae  4.0 \times 10^{15} \left(\frac{60}{N}\right)^{1/4}\mbox{ GeV}
 \;.
\end{equation}
However, from Eq. (\ref{6}), we can see that the curvature
perturbation produced from inflation will be much lower than the
observed value. It is fine though, if the curvature perturbation is
produced via a curvaton instead of the inflaton.
It is well known that introducing a curvaton
can reduce the scale of inflation (for example, see
\cite{Dimopoulos:2004yb, Dimopoulos:2002kt, Moroi:2005kz,Moroi:2005np}).

The method of introducing curvaton for dealing with the cosmic string
problem has been considered in \cite{Endo:2003fr, Rocher:2004my},
those papers are focused on dealing with the cosmic string
problems. On the other hand, conventional D-term inflation can be
ruled out if large non-Gaussianity is detected in future
experiment. In order to generate large non-Gaussianity, D-term
inflation have to be extended in some ways \cite{Bernardeau:2007xi}.
In this paper, we consider the possibility for generating large
non-gaussianity from a right-handed sneutrino field.

We assume that there is about $10\%$ contribution to the CMB power
spectrum coming from cosmic strings, which form after D-term hybrid
inflation which implies $H_\ast = 1.14 \times 10^{-7}M_P$. We are
interested in producing large non-gaussianity ($10<f_{NL}<100$).
From Eq. (\ref{eq15}) this implies $3 \times 10^{-6} <\Phi_\ast < 3
\times 10^{-5}$, this corresponds to $4.25 \times 10^{-10}<
\lambda_\nu <4.25 \times 10^{-9}$. For $m_\Phi=100\mbox{ GeV}$, from
Eq. (\ref{eq1}) the lightest neutrino mass is $10^{-10}$ eV. This
favors RH sneutrino leptogenesis \cite{McDonald:2004by}.

WMAP data prefers a red tilted spectrum with spectral index $n_s
\simeq 0.96$ \cite{Komatsu:2008hk} which is not easy to achieve in
the curvaton scenario. In our model, because we choose a small RH
sneutrino mass ($m_\Phi=100$ GeV), from Eq. (\ref{eq3}), the
spectral index is $n_s \simeq 1$. However there is a way out. For
$10\%$ cosmic string contribution to the CMB power spectrum. $n_s
\simeq 1$ is actually the required value \cite{Bevis:2007gh,
Battye:2006pk}.

\section{Conclusions}
\label{5}
The curvaton scenario allows to generate the observed level of density
perturbations with a lower scale of inflation. Thus when applied to
D-term hybrid inflation, cosmic string problem can be
evaded. Furthermore, in this paper, we shown that if we use
right-handed sneutrino with a small mass $100\mbox{ GeV}$ (which is
favored by RH sneutrino leptogenesis) as the curvaton, large
non-Gaussianity $f_{NL}=10-100$ can be produced.

\section*{Acknowledgement}
This work was supported in part by the NSC under grant No. NSC
96-2628-M-007-002-MY3, by the NCTS, the Boost Program of NTHU, and the
WCU program through the KOSEF funded by the MEST
(R31-2008-000-10057-0).

\end{document}